# Unformatted Digital Fiber-Optic Data Transmission for Radio Astronomy Front-Ends

Matthew A. Morgan
National Radio Astronomy Observatory, Charlottesville, VA 22903; matt.morgan@nrao.edu

J. Richard Fisher
National Radio Astronomy Observatory, Charlottesville, VA 22903; rfisher@nrao.edu

Jason J. Castro
National Radio Astronomy Observatory, Charlottesville, VA 22903; jcastro@nrao.edu

**ABSTRACT.** We report on the development of a prototype integrated receiver front-end that combines all conversions from RF to baseband, from analog to digital, and from copper to fiber into one compact assembly, with the necessary gain and stability suitable for radio astronomy applications. The emphasis in this article is on a novel digital data link over optical fiber which requires no formatting in the front-end, greatly reducing the complexity, bulk, and power consumption of digital electronics inside the antenna, facilitating its integration with the analog components, and minimizing the self-generated radio-frequency interference (RFI) which could leak into the signal path. Management of the serial data link is performed entirely in the back-end based on the statistical properties of signals with a strong random noise component. In this way, the full benefits of precision and stability afforded by conventional digital data transmission are realized with far less overhead at the focal plane of a radio telescope.

## 1. INTRODUCTION

Despite the widely recognized intrinsic value of electronic component integration, most high-performance radio astronomy front-ends today still retain the classical boundaries between analog, digital, and optoelectronic sub-assemblies, limiting the degree to which the potential benefits of system integration can ultimately be realized. While each of these three sub-systems may themselves be fully integrated, the interconnects that run between them remain as one of the dominant sources of mechanical failure, gain-frequency dependence, and gain instability in the entire receiver chain.

We report on the development of a fully integrated front-end comprising all non-cryogenic components of a radio astronomy receiver, taking as its input a broadband sky-frequency signal and delivering digital data at its output on optical fiber, with all the necessary signal transformation and conditioning in between (see Fig. 1). Successful integration of all these functions – analog, digital, and photonic – involves a number challenges. Prior publications on this work have focused on the analog-digital design, where the complementary natures of integration and digital signal processing (that is, stability and precision) are exploited to simplify front-end architecture while achieving high levels of performance (Morgan & Fisher 2010; Morgan, Fisher, & Boyd 2010; Morgan & Boyd 2011). Special attention was paid to the integration of analog and digital components in the front-end to ensure adequate isolation of the sample clock and its harmonics.

With this article, we extend that work by incorporating a novel digital fiber-optic transmission scheme which avoids formatting of the data in the front-end – that is, it does not use bit-scrambling, encoding, packetizing, or framing, nor does it transmit meta-data of any kind, only raw samples. This alleviates the potential problem of self-interference by minimizing the digital electronics present in the antenna. It also reduces the power consumption and bulk of the data transmission system, facilitating its integration with the analog components and the construction of large-format focal-plane arrays.

Any implementation of a high-speed, serial data transmission system must address the issues of DC balance, clock recovery, and word alignment. Without transmitter formatting, the approach taken in this work is to leverage the known properties of the data itself, which for a radio astronomy signal is well characterized by Gaussian-distributed white noise, even when strong interferers are present. In this way, the overhead for management of the serial data link is shifted entirely into the back-end. The specific manner in which each of these issues is dealt with is discussed in the following sections.

## 2. DC BALANCE

DC balance is needed because the components of a high-speed serial link are usually AC-coupled. Any DC offset is lost, resulting in a level shift in the eye diagram.

In the radio astronomy application, this is taken care of automatically. Individual samples are random with essentially a Gaussian probability distribution having zero mean value, and inspection of the sample codes in either two's-complement or straight/offset binary format, shown in Fig. 2, reveals that they are anti-symmetric about the mid-range; that is, positive sample codes are simply mirror images of the negative sample codes wherein the 1's have been replaced by 0's and vice-versa. These two facts together ensure that each bit in any given sample has an equal chance of being a 1 or a 0, and DC balance is achieved. All that is required is that the ADC's have reasonably low offset voltage. Minute offsets lead to correspondingly minute vertical shifts in the eye diagram, which are unlikely to break the serial link.

## 3. CLOCK RECOVERY

The conditions for successful clock recovery are somewhat more difficult to articulate, and different manufacturers will specify their requirements in different ways. Some quote the number of consecutive identical binary digits (presumably embedded in an otherwise pseudo-random bit sequence) which can be tolerated by the internal phase-locked loop, usually numbered in the thousands. Other manufacturers calculate their spec in terms of transition density, or the average number of bit changes that occur over a large number of clock cycles. In either case, conventional serial data links ensure that a sufficient number of transitions are present for clock recovery by employing bit scramblers.

For the radio astronomy application, the chances of raw samples (or "words") from a Gaussian noise source containing long strings of identical bits is vanishingly small. Transition density may be calculated by counting the number of transitions in each possible sample code and summing them, each with a weight given by the (Gaussian) probability of that sample code occurring,

$$p_n = \frac{1}{2} \begin{cases} \operatorname{erf}\left(\frac{(n+\frac{1}{2})v_0 - \mu}{\sigma\sqrt{2}}\right) + 1 & n = 0 \\ 1 - \operatorname{erf}\left(\frac{(n-\frac{1}{2})v_0 - \mu}{\sigma\sqrt{2}}\right) & n = 2^N - 1 \\ \operatorname{erf}\left(\frac{(n+\frac{1}{2})v_0 - \mu}{\sigma\sqrt{2}}\right) - \operatorname{erf}\left(\frac{(n-\frac{1}{2})v_0 - \mu}{\sigma\sqrt{2}}\right) & \text{otherwise} \end{cases} \quad (1)$$

where $n$ is the sample value, $v_0$ is the sampler threshold level, $\mu$ is the mean value of the signal relative to the bottom of the reference scale, and $\sigma$ is the standard deviation (rms, or root-mean-square amplitude).

For the boundaries between samples, it is assumed for the purposes of this analysis that the power spectrum of the noise is white over the Nyquist bandwidth, ensuring that adjacent samples are uncorrelated, although in practice this assumption can be considerably relaxed. For zero-mean noisy signals, this results in a bit transition at the word boundary 50% of the time. If the ADC has an offset, then it is necessary to consider the relative likelihood of sample codes in the upper and lower half of the range, respectively, in order to calculate the chances of a bit transition between words. Primarily, this leads to a perturbation in the limiting transition density at very low power levels.

The transition density as calculated in this manner is shown in Fig. 3. The dotted line at 50% is a benchmark for data streams having maximum entropy – that is, random sequences for which knowledge of one bit provides no extra information about what the next bit might be; it could be the same or different with equal probability. This would be typical of, say, a compressed data file. Higher transition density than this aids in clock recovery, but such data streams would be of diminishing value for efficient communication, as beyond this point they become more deterministic (100% transition density, for example, would correspond solely to the fixed sequence of alternating 1's and 0's). It is reasonable, therefore, to assume that commercial clock recovery circuits should at least be able to function with 50% transition density and probably a fair amount less. Indeed, some datasheets indicate that 40% is more than adequate.

It is evident from the plot that straight binary has an advantage over two's complement in terms of its transition density. This advantage becomes even more pronounced at lower bit resolutions.

The open circles indicate the power levels where optimum quantization efficiency would be achieved for the specified number of bits (Thompson et. al. 2007). It is seen that for both 4- and 8-bit sampling, this level corresponds approximately to the peak in transition density. That the optimal points fall close to the 50% maximum entropy line may not be a coincidence, as quantization efficiency cannot be independent of informational content.

For the present analysis, what matters most is that the nominal operating points lie well within a regime where clock recovery can be achieved without additional formatting. Moreover, if 40% transition density represents a safe lower bound for reliable operation, then the permissible operating range for both 4- and 8-bits is roughly 23 dB wide. (Later measurements suggest that 40% is much too conservative, and that the link can be made to work over a dynamic range that exceeds 50 dB.)

## 4. WORD ALIGNMENT

Upon successful recovery of the clock, the deserializer in the back-end module of Fig. 1 begins transferring the incoming data stream to its outputs in parallel format. However, the boundaries between individual sample words have been lost; the most-significant bit (MSB) from each sample now appears at a random location on the deserializer output pins. Word alignment is the process of detecting this boundary and then rotating the data bits on the output pins until the MSB appears in the correct location.

The mechanism for detecting the MSB is based on a statistical calculation of the correlation between adjacent bits. Inspection of the sample codes for straight binary encoding (see Fig. 2) reveals that in the entire middle half of the range the first and second most-significant bits are anti-correlated. The higher rate of occurrence of middle-half sample codes given by the Gaussian distribution ensures that these correlations will dominate in the data stream. The remaining bits have alternating correlation and are positively correlated in the most likely codes in the center. In fact, if one assumes that the analog signal is Gaussian-distributed with a white power spectrum, the exact probability of one bit being anti-correlated with its neighbor may be calculated,

$$q_k = \begin{cases} 0.5 & k = 0 \\ \sum_{i=1}^{2^{N-1-k}} (-1)^{i-1} \text{erf}\left(\frac{2^{N-1} - 2^k\left(i - \frac{1}{2}\right)}{\sigma\sqrt{2}} v_0\right) & 1 \leq k < N \end{cases} \quad (2)$$

where $k=0$ corresponds to the LSB and its adjacent bit is the MSB from the next uncorrelated sample (Morgan and Fisher 2009 & 2011).

The result of this calculation for different bits in the sample is plotted in Fig. 4. Note that the probability of anti-correlation for the MSB and the second most-significant bit is close to unity for all power levels except where the signal undergoes severe clipping at the rails (a condition which will be referred to as inversion, for reasons that will soon become clear). The same probability for any other bit pair is always 50% or less. This serves as a statistical marker that we can use to identify the MSB in the output words.

To do so, a single high-speed XOR gate is placed on the two most-significant pins of the deserializer. If the word is properly aligned, the output of the XOR gate (test point *A* in Fig. 1) should be 1 almost all the time. If not, the output is a 1 only 50% of the time or less. This waveform is then passed through an integrator or low-pass filter with a time constant extending over thousands or millions of sample periods, producing a DC voltage (test point *B*) which is directly proportional to the duty cycle of the XOR output. Finally, this voltage is checked against a pre-determined threshold by a comparator, resulting in a logic output that signals unambiguously when word alignment has been achieved. It is up to the link management circuitry (in the back-end) to trigger bit slips in the deserializer until word alignment is indicated.

Note that beyond a root-mean-square input signal amplitude of roughly 74% relative to the full-scale range of the sampler, the probability advantage of the MSB is lost and the algorithm fails. In Fig. 5, a comparison of the histograms at nominal power levels and at this crossover point helps to illustrate why this occurs. Recall that the algorithm depended on the predominance of sample codes in the middle half of the sampler range. At inversion, the outermost bins at the rails of the sampler have accumulated enough out-of-range samples that the integrated probability of the signal falling in the outer two quadrants begins to exceed 50% while that in the middle half drops below 50%. The probability distribution function has then inverted.

## 5. MEASUREMENTS

A prototype integrated receiver was developed and constructed according to Fig. 1 to verify these concepts. The front-end module is shown in Fig. 6. It has an RF frequency range of 1.2-1.7 GHz, with an instantaneous IF bandwidth of roughly 75 MHz per sideband, 150 MHz total. The ADC's have 8-bit resolution and are clocked at 155.5 MS/s by a crystal oscillator source integrated into the module. Both channels are connected to a 16-bit serializer driving a low-power Vertical Cavity Surface Emitting Laser (VCSEL). The fiber output is single-mode at 1310 nm wavelength and operating at 2.488 Gbps serial data rate.

At the other end of the fiber is an evaluation back-end consisting of an optical receiver board with built-in word alignment circuitry and a 16-bit deserializer (see Fig. 7). The parallel outputs are delivered by InfiniBand cable to a National Instruments PXI-6562 data acquisition card. Three visual LED monitors are provided on the back-end board for diagnostic purposes. The first LED lights when the photodiode is receiving an optical carrier. The second LED is lit when the deserializer indicates that the serial clock has been recovered. The final LED is tied to the output of the comparator that indicates when word alignment has been established. A pushbutton was provided to trigger the SYNC input on the deserializer (initiating a single bit rotation of the output word) but in practice it turned out to be simpler to drive this input automatically from the data acquisition computer.

The module was designed with appropriate gain for radio astronomy applications, assuming that it is driven by an external (usually cryogenic) amplifier. The initial tests, however, were conducted without a preamp. Instead a strong CW tone was injected directly into the RF port at 1 MHz offset from the local oscillator. In this way, sinusoidal IF tones were synthesized in the front-end with a noise floor almost too weak to be detected by the sampler. Although the word alignment algorithm was designed to operate on the Gaussian noise statistics that are typical of radio astronomy data, all that is really required is that the sample codes in the middle of the range are most frequent. Thus, a simple CW tone works well so long as its amplitude is not too large, and provides a means for easy visual inspection of the deserialized data for verification (bit transmission errors in essentially noisy data could very easily be missed.) The resulting waveforms are shown in Fig. 8. Some dithering is present which is attributed to residual noise in the front-end. The different amplitudes of the sinusoids in the I and Q channels resulting from gain mismatch is evident, as is the expected quadrature phase relationship between the two.

Following this, an RF preamp was added to the measurement setup in order to bring the noise floor up to nominal levels (in an actual radio astronomy front-end, this would correspond to the cryogenic gain inside the Dewar). At first, the RF input of the preamp was terminated so that the link may be tested with a noise-only input. Word-alignment was established, then data were recorded with the digital output words rotated through each of the sixteen possible positions to illustrate the appearance of the data in both aligned and unaligned conditions. The results are tabulated in Fig. 9. The first column describes the position of the MSB as determined by the word alignment indicator. The second and third columns are graphs of the time series and frequency spectra for one of the two output channels. Correct outputs are evident from the small amplitude of the time series, and low-pass shape of the power spectrum. Incorrect outputs appear much larger in amplitude and the passband shape is lost. Note that valid data appears in the first, ninth, and seventeenth row, corresponding to two complete word rotations with the channels swapping places once and then back again.

In the forth column is an oscilloscope trace taken at the output of the XOR gate (test point *A* from Fig. 1). With the words in proper alignment, this output is almost constant at the logic-high level. The single downward spike in the center of the trace on these rows corresponds to the rare logic-low output that triggers the oscilloscope. After a single bit-rotation, the trace changes to a predominately logic-low output with occasional logic-high bits. In all other cases, the duty cycle of the XOR output is on the order of 50%.

Finally, column five lists the DC voltage monitored at the output of the integrator (test point *B* from Fig. 1). In word-aligned position, the output is greater than 1 V. In all other positions, it is below 20 mV. The threshold for the comparator was set halfway between at 0.5 V.

Next, a CW tone was injected through the RF preamp in order to simulate the effects of a strong non-Gaussian source or interferer. Prior simulations had indicated that such interferers, even when equal in power to the total integrated noise in the band, only perturbed the probabilities in Fig. 4 slightly, and this only at the right edge of the plot where the sampler is near saturation (Morgan & Fisher 2009 and 2010). This test now confirms that result; nominal operation of the serial link is unaffected by such an input. The recorded time series and power spectra for both channels are shown in Fig.'s 10 and 11. In these tests the local oscillator was set to 1450 MHz, and the CW test tone to 1475 MHz.

## 6. OPERATIONAL LIMITS

The final tests of the serial link involved running at extreme analog power levels (both low and high) to determine when clock recovery and word alignment would fail, respectively.

As was indicated in Section 3, the transition density needed for reliable clock recovery may not be achieved at very low input power levels to the sampler, at which only the innermost two sample codes occur frequently. For straight binary encoding, the asymptotic transition density at infinitesimal signal power and assuming zero offset in the sampler is given by $1.5/N$, where $N$ is the number of bits per sample. Interestingly, for samplers with small offsets the limiting transition density actually improves to $2/N$. Thus, the minimum low-power transition density for 8-bit sampling in straight binary is somewhere between 18-25%. For fewer numbers of bits it is much larger.

In practice, one should expect clock recovery failure due to low transition density to be a gradual phenomenon, wherein momentary lapses in clock-to-data synchronization cause bits to be lost or corrupted with increasing regularity. An experiment was performed wherein the input noise power to the prototype module was gradually reduced by adjusting the preamp bias until the clock recovery loop failed momentarily, allowing a bit to slip. A snapshot of the data was recorded at each power level and the transitions counted in order to verify the theory, as shown in Fig. 12. (Minor deviations from the theoretical curve may result from mis-estimation of the preamp's gain at low bias levels.) The tests showed that an rms amplitude much less than a single sampler threshold (less than 20% transition density in this configuration) was required for the bit errors to occur frequently enough that one could be observed in the 5-minute time scale of the experiment – which, at 155.5 MS/s, corresponds to a slip rate of approximately 1 out of every $10^{10}$ bits.

Importantly, further reduction in power does not make the bit slips any more frequent than this – by that point the sampler is only detecting zero-crossings and the statistics relating to transition density cease to change (if anything, the transition density would be on the rise again as a result of the sampler offset phenomenon discussed above.) Also, bit *values* were not being appreciably corrupted at this point either, since that would tend to produce erratic sample codes well outside the vanishingly small rms of the signal, and no such outliers were observed.

It is interesting to note that while the clock recovery circuit may fail intermittently at extremely low power levels and large bit resolutions, manifesting as the infrequent loss of one or more bits, Fig. 4 shows that the word alignment algorithm should continue to work perfectly even down to infinitesimal power levels. The system is therefore automatically capable of detecting when it has missed a bit and operation may continue in this state so long as it is designed to monitor the status and re-align the word boundaries when needed.

The final test was to increase the analog power level until word alignment failed due to an excess of outer-quadrant sample codes. Rather than further amplifying the noise power, the amplitude of the CW tone was increased to simulate the effects of a very strong interferer with non-Gaussian characteristics – a far more likely scenario in practice. The tests showed that with a nominal noise amplitude of 15 levels rms, a CW power 14 times stronger than the noise was required to invert the spectrum. The final histogram at this break-point is shown in Fig. 13, along with a theoretical model consisting of the convolution of histograms for Gaussian and sinusoidal waveforms, respectively. Asymmetry in the measured histogram is probably the result of uneven compression in the analog components prior to digitization.

Note that inversion of the probability density function in this configuration occurs sooner than it would have for a pure noise signal according to Fig. 4. This is due to the U-shaped histogram of a sinusoidal waveform. A strong CW tone exhibits a predominance of outer-quadrant sample codes at much lower power levels than pure Gaussian noise. Still, the presence of a CW tone 14 times more powerful than the integrated noise in the IF band is highly unlikely in practice. Interestingly, adding more noise would likely improve the reliability at these high power levels as it would tend to restore the expected Gaussian profile to the histogram.

## 7. FUTURE WORK

In this implementation of the receiver front-end, two analog channels have been fed into the same serializer, resulting in the samples from each channel being interleaved within the data stream. This could be a common configuration in real implementations, as many commercial serializers operate on 16-bit words, and there is rarely any reason for radio astronomy data to be digitized at such a resolution. The word alignment algorithm described herein has no way of determining which half of the 16-bit word corresponds to the first channel and which corresponds to the second, only that the first pin is the MSB of one of them.

There are a number of ways this could be handled at a higher level using operational protocols. One is to note that the analog outputs correspond to in-phase and quadrature-channels of a sideband separating mixer, so a calibration signal (or other single-sideband observational source) should have close to the expected +90 degree phase correlation in the two channels; if it has the opposite sign then the output words need to be swapped. Alternatively, since the net gains of the independent analog channels will almost certainly differ by a couple of dB, and the phase relationship is to be calibrated post-construction anyway (Morgan & Fisher 2010) one may simply adopt the convention that the channel with the strongest amplitude will always appear first on the deserializer output.

Also, although it was not attempted in this proof-of-concept, many instrument architectures are likely to include several links such as this in parallel corresponding to different polarizations or different beams in a phased array, where it may be important to track the relative propagation delay between fiber channels. These delays may drift during operation as a result of flexure on the cable wraps at the telescope axes, or by thermal expansion over long distances.

This issue is quite familiar to engineers developing analog fiber-optic links, and the solutions that might be employed are quite similar, with one important exception: The differential delay between any two recovered channels in this scheme is discrete, corresponding to an integer multiple of the sample length. In cross-correlation, the relative phase between two channels would have a discrete linear slope with respect to frequency, the possible values of which may be predicted a-priori. Therefore, one needs only to monitor and track the delay on separate fibers to within a sample period, after which the correction for fiber-optic delays will be exact.

Finally, the net power dissipation of all the non-cryogenic analog, digital, and photonic components of a radio astronomy front-end is quite substantial when integrated into a single compact module such as this. The current implementation draws 2.2 W of power – reasonably efficient for a complete analog-digital-photonic subsystem, but further reduction would be of great value in building large focal plane arrays. Projections for higher-speed (10 Gbps) implementations are that they would require roughly 3.5-4.0 W of power, with almost half being dissipated in the ADCs, and another quarter in the serializer.

The interface between the ADCs and the serializer is an area where improvements could be made. Since the direct connection of a very high-speed ADC and serializer is unique to our design, the interface between these two components using off-the-shelf parts is poorly optimized. Anticipating that the samples from a high-speed ADC would be clocked by some form of complex logic device or micro-processor, the ADC outputs are typically delivered off-chip using power hungry, transmission-line based, differential logic protocols, such as Low-Voltage Differential Signaling (LVDS). These bits are then gathered by the serializer (in our application) using resistively terminated LVDS receivers to be reformatted serially for transfer to the laser driver, as shown in Fig. 14a. This entire process is wasteful of power, pins, and circuit board real estate. A better solution, shown in Fig. 14b, would be to integrate both functions on to the same chip, avoiding the need for off-chip resistor-loaded transmission-lines. It is estimated that roughly a third of the ADC's power dissipation and 10-15% of the total power budgeted for the front-end could be saved at this interface alone. (Note that the intent is for the serial output to run continuously in real-time at fiber-link serial data rates, say 2.5 Gbps or 10 Gbps, distinguishing it from other serial-output ADCs which either buffer their data for slower readout or deliver it on serial LVDS lines at lower speeds along with synchronous serial- and frame-rate clocks for immediate latching into a data processor.)

It is also true that lower bit resolution, which is acceptable in radio astronomy, could offer significant power savings. Industrial applications, however, generally push for greater bit resolution, limiting the availability of lower bit resolution components. The lack of a parallel interface in the proposed topology of Fig. 14b opens the door for variable-resolution ADC architectures.

## CONCLUSION

We have presented a complete, non-cryogenic front-end receiver assembly for radio astronomy applications with all the conversions from RF to baseband, from analog to digital, and from copper to fiber in a single integrated module. Key to the success of this module is a novel unformatted fiber-optic link which minimizes the amount of digital hardware required at the focal plane of the telescope, reducing bulk and power dissipation while mitigating the potential self-generated RFI. This enables large-format focal-plane arrays to be constructed with in-situ digitization, helping to confine the analog amplitude and phase drifts to the integrated receiver where they are smallest, for improved calibration longevity and performance.

Although the principles of operation are based on Gaussian white noise, tests on the serial data link show that it is robust and reliable in the presence of strong interferers and real-world passbands. The implementations of both front-end and back-end are simple and readily realizable using off-the-shelf components (ADCs, serializers, deserialiers, and high-speed logic gates). It is also scalable; the prototype reported operates at a serial rate of 2.5 Gbps, but extension to 10 Gbps is straightforward using available components. Reduced power dissipation and wider analog bandwidth commensurate with modern radio astronomy requirements may be achieved using lower bit resolutions, for which the link is even better suited (due to enhanced transition density at low power levels.)

## ACKNOWLEDGEMENTS

The authors would like to thank Bill Shillue, Francoise Johnson, and Christophe Jacques for their help in troubleshooting various aspects of this prototype. The National Radio Astronomy Observatory is a facility of the National Science Foundation operated under cooperative agreement by Associated Universities, Inc.

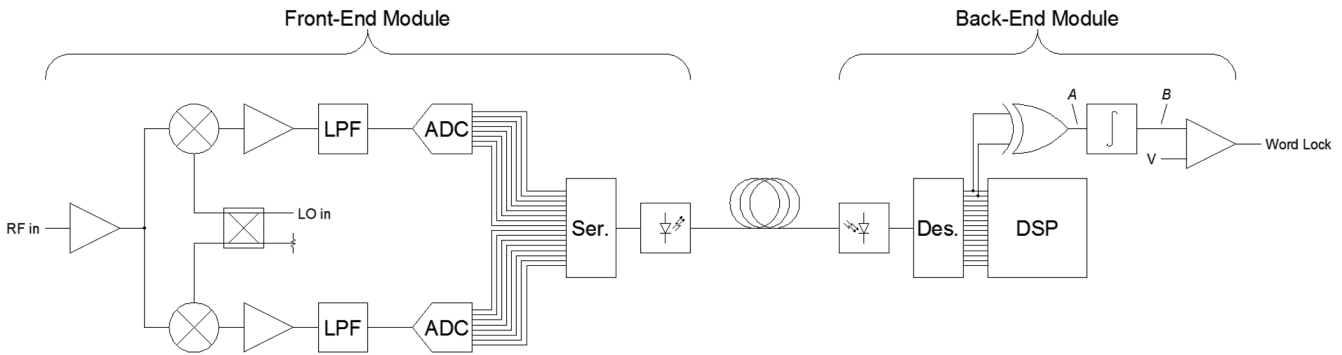

Fig. 1. Simplified block diagram of the prototype receiver using an I/Q mixer pair, including integrated analog-digital-photonic front-end module and evaluation back-end with link-management components.

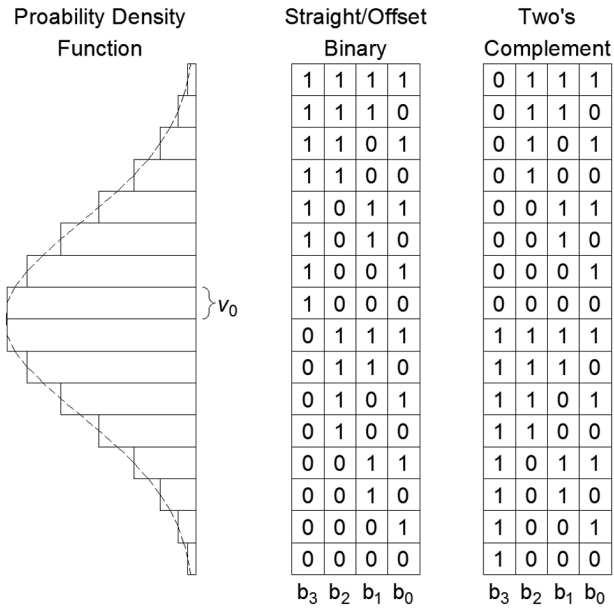

Fig. 2. Straight Binary and Two's Complement encoding of Gaussian distributed signals.

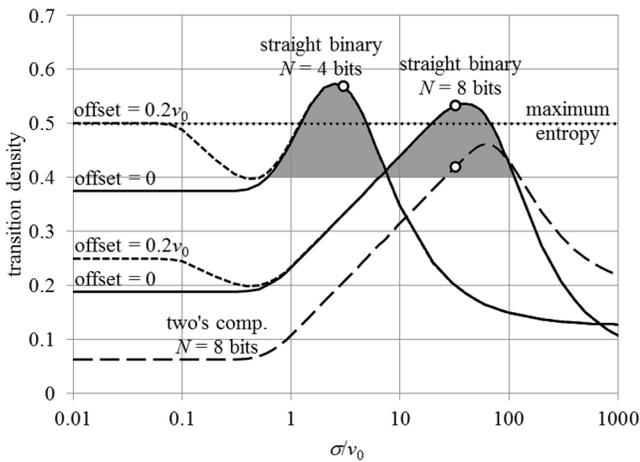

Fig. 3. Serial data transition density versus signal amplitude. The analog noise source is assumed to be Gaussian distributed with a white power spectrum. Open circles represent levels that achieve optimum quantization efficiency, and the shaded regions represent nominal operating ranges.

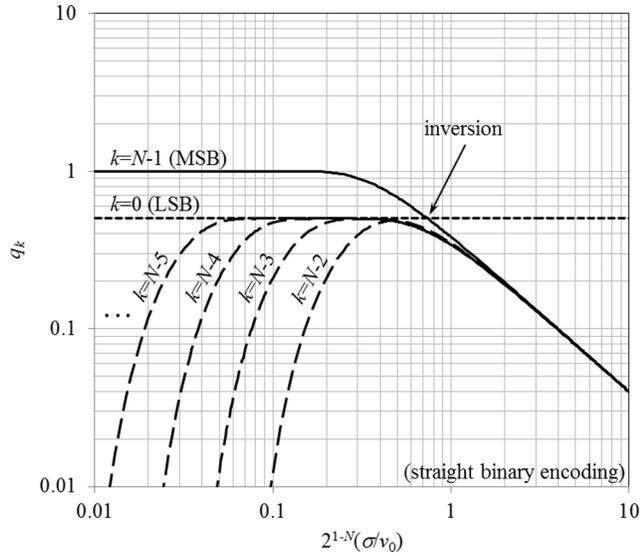

Fig. 4. Probability ($q_k$) that bit $k$ does not equal bit $k$-1, where $k$=0 corresponds to the least significant bit (LSB) and $k$=$N$-1 corresponds to the most significant bit (MSB). This result is independent of bit-resolution.

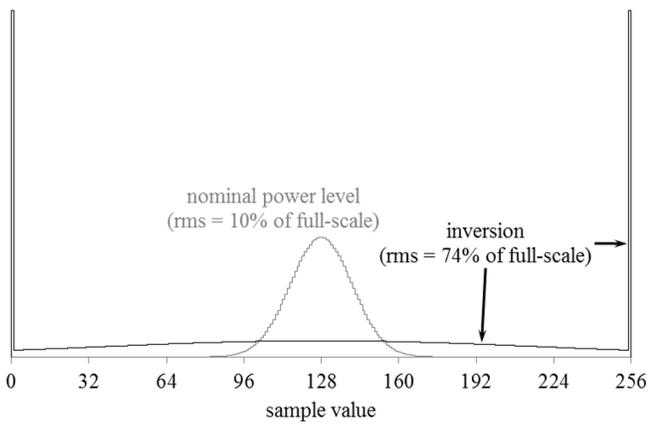

Fig. 5. Comparison of sampler histograms where the signal power is either nominal or at inversion. Data shown for 8-bit resolution.

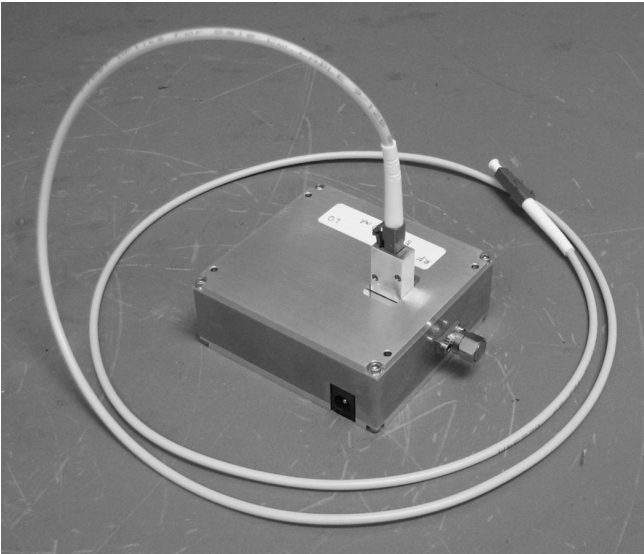

Fig. 6. Integrated analog-digital-photonic receiver front-end. Dimensions are 2.8" x 2.6" x 1.0".

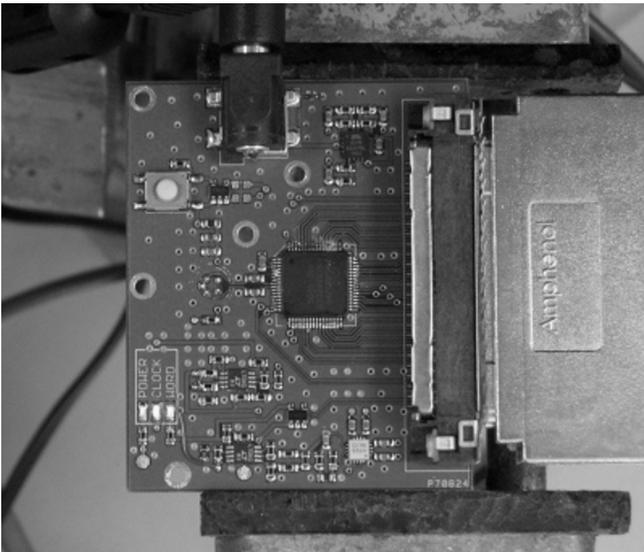

Fig. 7. Evaluation back-end consisting of optical receiver with word alignment circuitry. The large IC in the center is the deserializer. Board dimensions are 2.0" x 2.4".

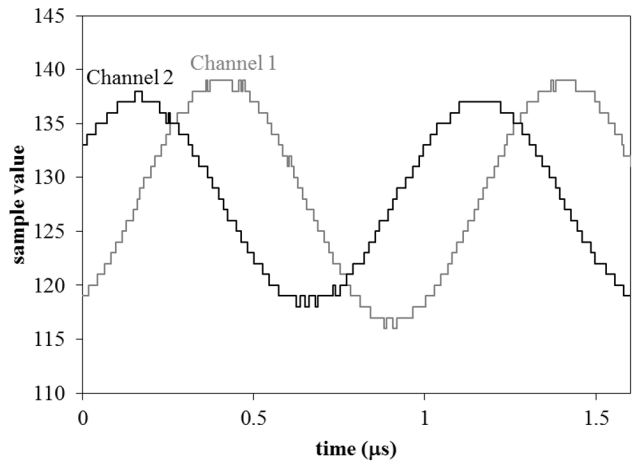

Fig. 8. Quadrature sinusoid test signals after transmission over unformatted fiber optic link. Dithering is due to residual noise in the front-end.

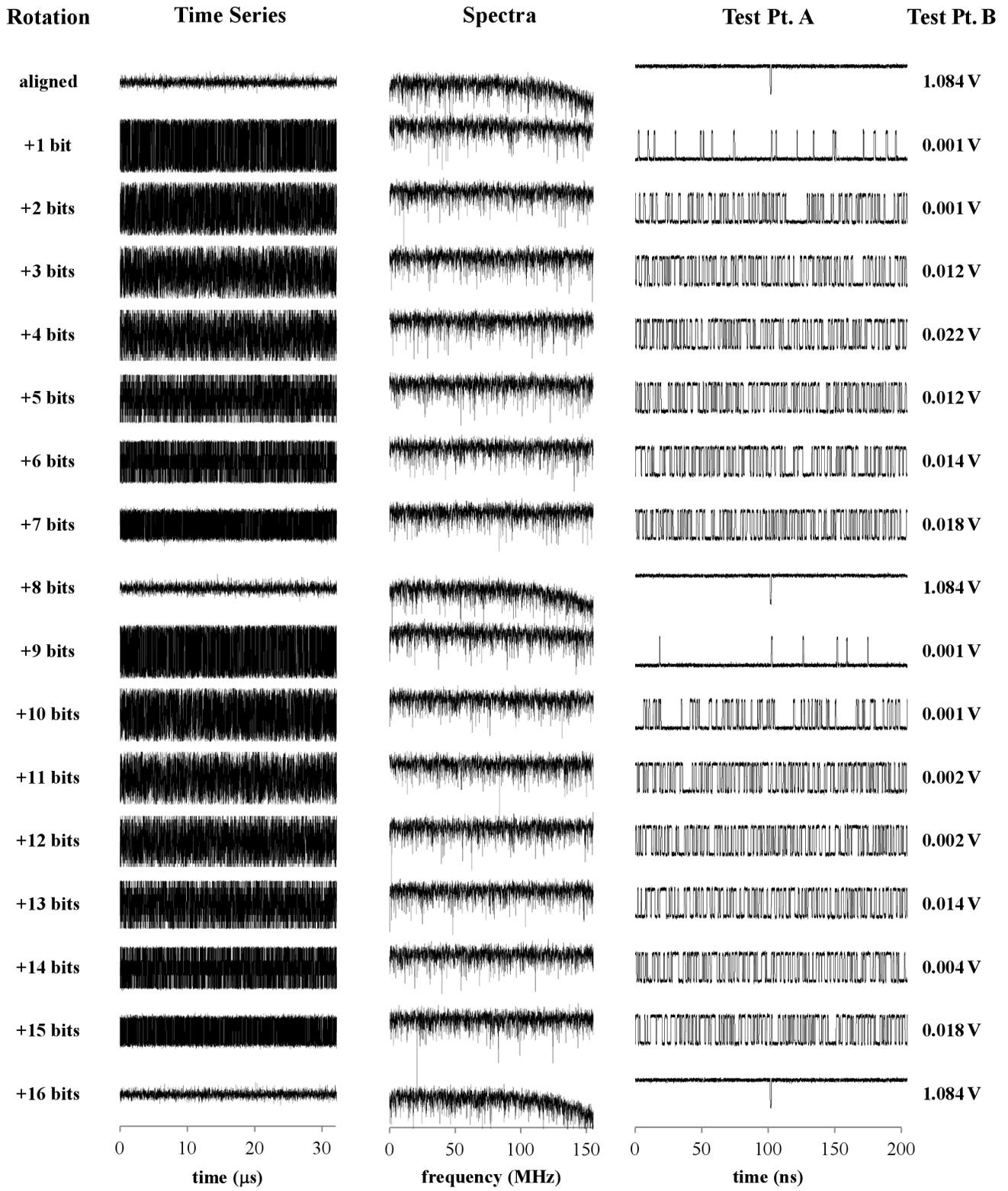

Fig. 9. Output data for unformatted fiber optic link with broadband noise input.

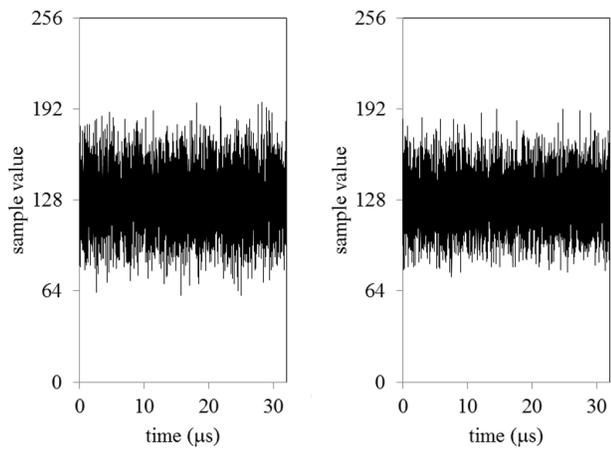

Fig. 10. Time-series for both output channels with broadband noise and CW tone input.

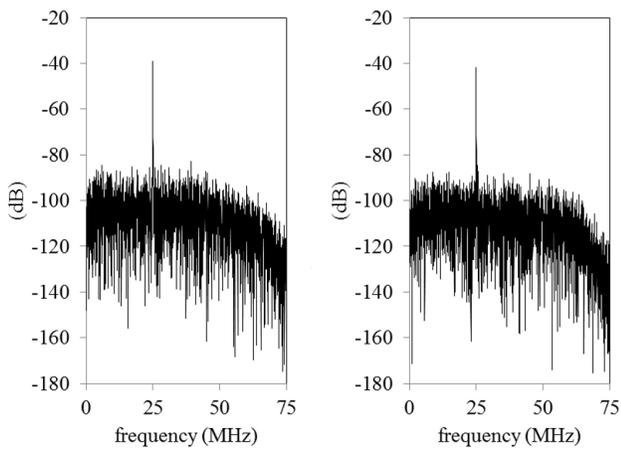

Fig. 11. Power spectra for both output channels with broadband noise and CW tone input.

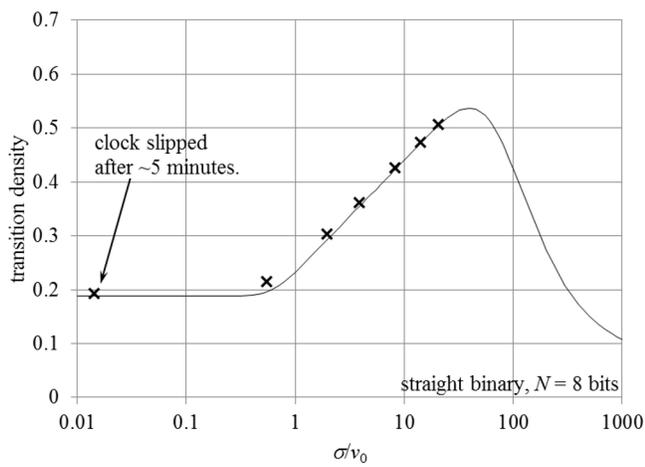

Fig. 12. Measured (markers) and modeled (solid line) transition density versus signal amplitude. The clock recovery loop failed momentarily after roughly 5 minutes when the transition density dropped below 20%.

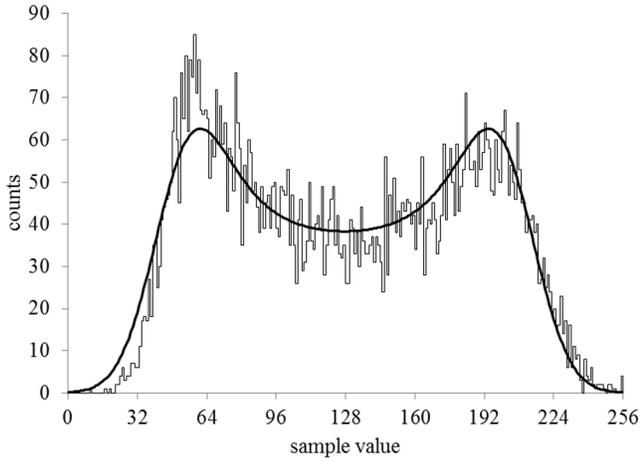

Fig. 13. Measured and modeled histogram with large-amplitude CW tone superimposed on broadband noise at the point where the word alignment algorithm fails.

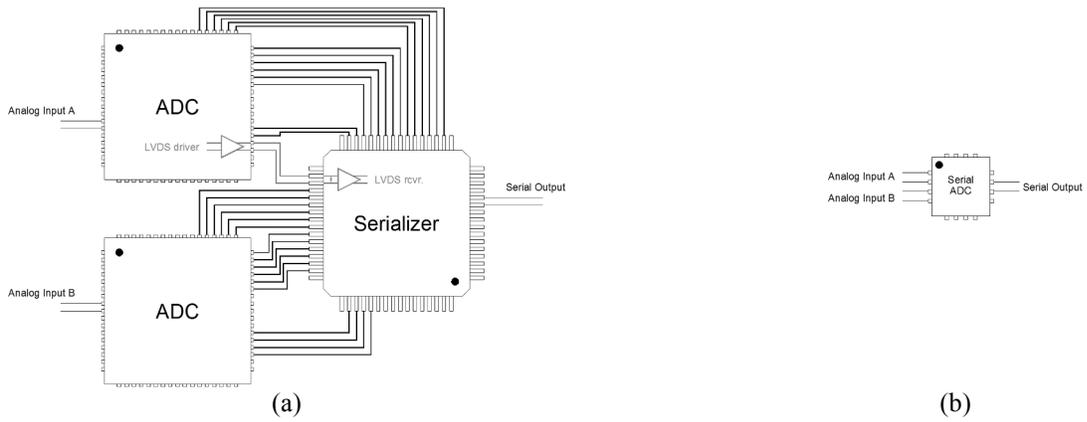

Fig. 14. Illustration of (a) conventional cascaded ADC and Serializer with LVDS ports, and (b) proposed combined ADC / Serializer with variable bit-resolution.